\documentstyle[12pt]{article}
\newcommand{\QED}{\rule{0.4em}{2ex}}
\begin{document}
\def\theorem#1#2{\vskip 8pt{\noindent\bf #1}{\it #2}\vskip 8pt\noindent}
\def\head#1{{\normalsize\bf #1}}


\begin{center}
{\large {\bf GLOBAL EXISTENCE FOR WAVE MAPS\\
WITH TORSION}}
\vskip 1.5cm
{\bf  Stephen C. Anco${}^1$} and {\bf  James Isenberg${}^2$}
\vskip .75cm
{${}^1$ Department of Mathematics\\
Brock University, St Catharines, ON L2S 3A1, Canada\\
Email: sanco@brocku.ca}
\vskip.5cm
{${}^2$ Department of Mathematics and Institute of Theoretical Science\\
University of Oregon, Eugene, OR 97403-5203, USA\\
Email: jim@newton.uoregon.edu}
\end{center}

\begin{abstract}
Wave maps (i.e. nonlinear sigma models) with torsion
are considered in 2+1 dimensions.
Global existence of smooth solutions to the Cauchy problem
is proven for certain reductions under a translation group action:
invariant wave maps into general targets,
and equivariant wave maps into Lie group targets.
In the case of Lie group targets (i.e. chiral models),
a geometrical characterization of invariant and equivariant
wave maps is given in terms of a formulation using frames.
\end{abstract}

\section{\head{INTRODUCTION}}

There has been considerable progress during the last ten years
in the mathematical study of long-time existence properties of solutions of
geometrically-based classical field theories.
A significant portion of this work has focussed on the study of
what are called wave maps in the mathematics literature
and nonlinear sigma models (or, in certain special cases, chiral models)
in the physics literature.
These are defined as maps $\psi$ from a Lorentzian geometry
$(M^{m+1},\eta)$, e.g. Minkowski space,
to a Riemannian geometry $(N^{n},g)$, e.g. a symmetric space
or a compact Lie group,
with $\psi$ being a critical point for the functional
\footnote{ Integrals over $M^{m+1}$ are understood to use
the natural volume form compatible with the metric $\eta$.}
\begin{eqnarray}
S[\psi] = \int_{M^{m+1}}
\eta^{\mu\nu} g_{AB}(\psi) \partial_\mu\psi^A \partial_\nu\psi^B,
\label{S}
\end{eqnarray}
and hence satisfying the wave map equation
\begin{eqnarray}
\eta^{\mu\nu}\nabla_\mu \partial_\nu \psi^A +
\Gamma^A{}_{BC}(\psi) \partial_\mu\psi^B \partial_\nu\psi^C \eta^{\mu\nu}
=0 ;
\label{wavemapeq}
\end{eqnarray}
here $\nabla$ is the (torsion-free) derivative operator
determined by the metric $\eta$,
and $\Gamma$ are the (torsion-free) connection coefficients
compatible with $g$.

Wave maps have a well-posed Cauchy problem,
and it is known that for 1+1 dimensional base geometries
$(M^{1+1},\eta)$,
every choice of smooth initial data evolves into a global smooth solution
\cite{Gu,GenibreVelo,Shatah},
while for 3+1 (or higher) dimensional base geometries,
certain smooth initial data leads to solutions with singularities
\cite{Shatah,ShatahT-Z}.
Not yet understood is what happens in general for
2+1 dimensional base geometries.
This is the ``critical dimension'' (see \cite{Klainerman,Shatah-Struwe}),
where global smooth solutions are expected, at least,
for all smooth initial data of sufficiently small energy.
While global existence results are known to hold
for certain classes of rotationally-symmetric wave maps
in 2+1 dimensions (without restrictions on the energy)
\cite{ChristodoulouT-Z,ShatahT-Z}, not much is known otherwise for
critical wave maps \cite{Shatah-Struwe}.

An interesting modification to the wave map equations can be
obtained by adding torsion. This can be done in 2+1 dimensions,
without adding extra dynamical fields, as follows. 
One fixes a pair of
background fields: a closed one-form field $v$ on the base manifold
$M^{2+1}$ and a non-closed two-form field $p$ on the target $N^n$.
The field $p$ serves as a ``torsion potential'' in the sense that the
torsion tensor on $N^n$ is defined as
\begin{eqnarray}
 Q^A{}_{BC} = 3/2 g^{AD} \partial_{[D} p_{BC]} .
\label{Q}
\end{eqnarray}
A map $\psi:M^{2+1} \rightarrow N^n$ is defined to be a torsion wave map
if it is a critical point for the functional
\begin{eqnarray}
 S_{\rm tor}[\psi] = \int_{M^{2+1}}  \Big(
\eta^{\mu\nu} g_{AB}(\psi) \partial_\mu\psi^A \partial_\nu\psi^B
+\lambda \epsilon^{\mu\nu\sigma} v_\sigma p_{AB}(\psi)
\partial_\mu\psi^A \partial_\nu\psi^B  \Big)
\label{Stor}
\end{eqnarray}
where $\lambda$ is a coupling constant
and $\epsilon$ is the 2+1 volume tensor
normalized with respect to $\eta$.
The torsion wave map equation obtained from (\ref{Stor})
is given by
\begin{eqnarray}
 \eta^{\mu\nu}\nabla_\mu \partial_\nu \psi^A +
{\tilde\Gamma}^A{}_{BC}(\psi) \partial_\mu\psi^B \partial_\nu\psi^C
\Big( \eta^{\mu\nu} +\lambda \epsilon^{\mu\nu\sigma} v_\sigma \Big)
=0
\label{torwavemapeq}
\end{eqnarray}
where
\begin{eqnarray}
{\tilde\Gamma}^A{}_{BC} = \Gamma^A{}_{BC} + Q^A{}_{BC}
\label{torsion}
\end{eqnarray} are the connection coefficients 
\footnote{The contorsion coefficients ${\tilde\Gamma}^A{}_{[BC]}$
compatible with $g$ as determined by the torsion
are identically equal to $Q$.}
compatible with $g$,
with torsion $Q$.
Note that the effect of the torsion is to add the nonlinear term
\begin{eqnarray}
\lambda \epsilon^{\mu\nu\sigma} v_\sigma
Q^A{}_{BC}(\psi) \partial_\mu\psi^B \partial_\nu\psi^C
\label{torsionterm}
\end{eqnarray}
to the wave map equation (\ref{wavemapeq}).

Wave maps without torsion have a conserved,
symmetric stress-energy tensor \cite{Klainerman}
arising from the functional $S[\psi]$.
With the addition of torsion,
the corresponding symmetric stress-energy tensor
obtained from the functional $S_{\rm tor}[\psi]$
is no longer conserved. 
However, we point out that a non-symmetric stress-energy tensor
can be derived
by considering the variation of $S_{\rm tor}[\psi]$
under infinitesimal diffeomorphisms of $M^{2+1}$ 
acting on $\eta$, $\epsilon$, $v$, and $\psi$.
This leads to
\begin{eqnarray}
T^\mu{}_\alpha =
\eta^{\mu\nu} g_{AB} \partial_\nu\psi^A \partial_\alpha\psi^B
-1/2 \delta^\mu{}_\alpha \eta^{\nu\sigma} g_{AB}
\partial_\nu\psi^A \partial_\sigma\psi^B\nonumber \\
+1/2 \lambda \epsilon^{\mu\nu\sigma} v_\alpha p_{AB}
\partial_\nu\psi^A \partial_\sigma\psi^B
\label{T}
\end{eqnarray}
which satisfies
\begin{eqnarray}
\nabla_\mu T^\mu{}_\alpha =
1/2 \lambda \epsilon^{\mu\nu\sigma}
\partial_\mu \psi^A \partial_\nu \psi^B p_{AB} \nabla_\alpha v_\sigma .
\end{eqnarray}
Hence $T^\mu{}_\alpha$ is conserved if $v$ is covariantly constant
on $(M^{2+1},\eta)$.
Furthermore, $T^\mu{}_\alpha$
reduces to the standard symmetric stress-energy tensor
for wave maps without torsion when $v$ is set to zero.
The stress-energy tensor (\ref{T}) is central to investigating
global existence for torsion wave maps.

The critical dimension for torsion wave maps,
just as for standard wave maps, is 2+1.
While we do not attempt here to investigate the general class
of critical torsion wave maps, we are able to prove global
existence for various reductions of critical wave maps,
with and without torsion,
where the base geometry is Minkowski space.
These reductions are defined by
the invariance or equivariance of the wave map $\psi$
under a one-dimensional group of translations acting on $M^{2+1}$.
More specifically,
choose Cartesian coordinates $(x, y, t)$ for $(M^{2+1},\eta)$
and denote the translation group action by
$(x, y, t) \rightarrow (x, y + \lambda, t)$.
Then, for any target $N^n$,
a wave map $\psi$ is {\it translation invariant} if
\begin{eqnarray}
\psi^A(x, y + \lambda, t) = \psi^A(x, y, t) . 
\label{transinvar}
\end{eqnarray}
Translation equivariant wave maps require that
the target $N^n$ admit a translation group action.
Let $\rho^A{}_B(\lambda)$ denote a representation of
the translation group action on the base $M^{2+1}$
acting on the target $N^n$.
Then a wave map $\psi$ is {\it translation equivariant} if
\begin{eqnarray}
\psi^A(x, y + \lambda, t) = \rho^A{}_B(\lambda) \psi^B(x, y, t) . 
\label{transequivar}
\end{eqnarray}
Note that translation equivariance (\ref{transequivar})
reduces to translation invariance (\ref{transinvar})
when (and only when) the representation $\rho(\lambda)$
is chosen to be trivial, $\rho^A{}_B(\lambda)= \delta^A{}_B$.

One class of targets for which there is a natural
translation group action available are Lie groups, $G$.
For a Lie group target $N^n=G$,
left and right multiplication on $G$ by
a one-parameter exponential subgroup $\exp(\lambda A)$
define translation group actions,
where $A$ is any element in the Lie algebra of $G$.
This leads to three types of equivariance as follows.
Let $\Psi$ devote a matrix representation of the wave map
$\psi: M^{2+1} \rightarrow G$
and let $L$ and $R$ be matrix representations of
elements of the Lie algebra of $G$.
Then $\psi$ is said to be, respectively,
{\it left-translation equivariant } if
\begin{eqnarray}
\Psi(x, y + \lambda, t) = \exp(\lambda L)\Psi(x, y, t) , 
\label{leftequivar}
\end{eqnarray}
or {\it right-translation equivariant } if
\begin{eqnarray}
\Psi(x, y + \lambda, t) = \Psi(x, y, t)\exp(\lambda R) , 
\label{rightequivar}
\end{eqnarray}
or {\it conjugate-translation equivariant } if
\begin{eqnarray}
\Psi(x, y + \lambda, t) = \exp(\lambda L)\Psi(x, y,t)\exp(\lambda R) . 
\label{conjequivar}
\end{eqnarray}

Corresponding to invariant wave maps (\ref{transinvar})
and equivariant wave maps
(\ref{leftequivar}), (\ref{rightequivar}), (\ref{conjequivar}),
we have the following four classes of reductions:\\

\noindent{\bf Invariant Wave maps} (Any target)
\begin{eqnarray}
\psi = \phi(x,t)
\label{invarreduce}
\end{eqnarray}
\noindent{\bf Left-Equivariant Wave maps} (Lie group target)
\begin{eqnarray}
\Psi = \exp(yL) \Phi_L(x,t)
\label{leftreduce}
\end{eqnarray}
\noindent{\bf Right-Equivariant Wave maps} (Lie group target)
\begin{eqnarray}
\Psi = \Phi_R(x,t)\exp(yR)
\label{rightreduce}
\end{eqnarray}
\noindent{\bf Conjugate-Equivariant Wave maps} (Lie group target)
\begin{eqnarray}
\Psi = \exp(yL)\Phi_C(x,t)\exp(yR)
\label{conjreduce}
\end{eqnarray}
In each case the 2+1 wave map equation for $\psi$
yields a 1+1 reduced equation
for $\phi,\Phi_L,\Phi_R,\Phi_C$, respectively,
provided that the target geometry is suitably invariant
as discussed later. 

We establish global existence of solutions
to the Cauchy problem for the class of translation-invariant wave maps
with torsion in Section 2.
While the proof for these wave maps is very similar
to that for $1 + 1$ wave maps with no torsion,
the torsion terms do introduce some subtleties into the analysis,
which we highlight.

In order to prove global existence of solutions to the Cauchy problem
for the three classes of translation equivariant wave maps with torsion,
we find it useful to work with a frame formulation
for 2 + 1 wave maps .
In Section 3 we introduce the frame formulation for general targets
and then proceed to relate wave map equivariance
for Lie group targets to frame invariance and equivariance.
In particular, our global existence theorems for equivariant wave maps
have a natural formulation and proof using frames.

The proof for the
left equivariant, right equivariant, and conjugate equivariant wave maps
with torsion is fairly similar in each case.
We focus on the left equivariant case
(which corresponds to invariant frames)
and carry out the global existence proof in detail, in Section 4.
We then briefly note in Section 5 the differences
entailed in proving global existence for the other two cases.
We make a few concluding remarks in Section 6.

\section{\head{INVARIANT WAVE MAPS WITH TORSION}}

The translation invariance condition (\ref{transinvar}) is characterized
by the wave map functions (\ref{invarreduce}) being independent of $y$.
Under this reduction
the torsion wave map equation (\ref{torwavemapeq}) becomes
\begin{eqnarray}
\gamma^{\alpha \beta} \partial_{\alpha} \partial_{\beta}\phi^A
+\tilde{\Gamma}^A_{BC} (\phi) \partial_{\alpha}\phi^B
\partial_{\beta}\phi^C
(\gamma^{\alpha \beta} + \lambda v_y \epsilon^{\alpha \beta})
= 0
\label{new14}
\end{eqnarray}
where $\gamma^{\alpha \beta}$ is the 1 + 1 Minkowski metric
($\alpha$, $\beta$ run over $x$ and $t$)
and $\epsilon^{\alpha \beta}$ is the 1 + 1 Levi-Civita tensor.
We hereafter take $v_y$ to be constant,
but we make no further restrictions:
The target $(N^n, g)$ can be any Riemannian geometry,
and the torsion potential $p$ can be any non-closed two form on $N^n$.

Interestingly, while the torsion term
\begin{eqnarray}
\lambda v_y
Q^A{}_{BC}(\phi) \partial_{\alpha}\phi^B \partial_{\beta}\phi^C
\epsilon^{\alpha \beta}
\label{new15}
\end{eqnarray}
appears in a nontrivial way in the reduced wave map equation
(\ref{new14}), and while the stress-energy tensor (\ref{T})
generally contains a torsion term,
for translation invariant wave maps
the torsion drops out of many of the stress-energy tensor components. 
We have
\begin{eqnarray}
T_{tt} = T_{xx} &=&
{1 \over 2} (|\partial_t \phi|^2 + |\partial_x \phi|^2) , 
\label{new16}
\end{eqnarray}
\begin{eqnarray}
T_{xt} = T_{tx} &=&
\partial_t \phi^A \partial_x \phi^B g_{AB} , 
\label{new18}
\end{eqnarray}
all of which contain no torsion, along with
\begin{eqnarray}
T_{yy} &=&
{1 \over 2} (|\partial_t \phi|^2 -|\partial_x \phi|^2)
+ {1 \over 2} \lambda v_y\epsilon^{\alpha \beta}
\partial_{\alpha}\phi^A \partial_{\beta} \phi^B p_{AB} , 
\label{new20}
\end{eqnarray}
\begin{eqnarray}
T_{xy} &=& 0 , 
\label{new21}
\end{eqnarray}
\begin{eqnarray}
T_{yx} &=&
{1 \over 2} \lambda v_x \epsilon^{\alpha \beta} 
\partial_{\alpha}\phi^A \partial_{\beta} \phi^B p_{AB} , 
\label{new22}
\end{eqnarray}
\begin{eqnarray}
T_{ty} &=& 0 , 
\label{new23}
\end{eqnarray}
\begin{eqnarray}
T_{yt} &=&
{1 \over 2} \lambda v_t \epsilon^{\alpha \beta} 
\partial_{\alpha}\phi^A \partial_{\beta} \phi^B p_{AB} . 
\label{new24}
\end{eqnarray}
Note that $v_x$ and $v_t$ do not appear in
the reduced wave map equation (\ref{new14});
setting them to zero does not affect (\ref{new14}),
but it does result in $T_{yx}$ and $T_{yt}$ vanishing.

We now consider the Cauchy problem for
translation invariant wave maps (\ref{invarreduce})
with torsion.
Initial data at $t=t_0$ consists of
a pair of maps
\begin{eqnarray}
\hat{\phi}: \Sigma \rightarrow N, \qquad
\hat{\theta}:\Sigma \rightarrow TN
\label{new25}
\end{eqnarray}
(here $\Sigma = R^1$ or $S^1$ allowing for periodic boundary conditions).
A solution to the Cauchy problem is then a map $\phi: \Sigma \times
R^1
\simeq M^{2+1}
\rightarrow N$  which satisfies (\ref{new14}) along with the initial
conditions
\begin{eqnarray}
\phi(x, t_0) = \hat{\phi}(x), \qquad
\partial_t \phi (x, t_0) = \hat{\theta}(x) . 
\label{new26}
\end{eqnarray}
Note that there are no constraints on the choice of initial
data $\{\hat{\phi}, \, \hat{\theta}\}$.
Global existence of initial value solutions
is established by the following theorem.

\begin{theorem}{Theorem 1.}{
For any smooth compact support
{\footnote{  $\hat{\phi}$ is compactly supported
if it is constant everywhere outside a compact region in $\Sigma$;
$\hat{\theta}$ is compactly supported if it zero outside such a region.} }
initial data,
the Cauchy problem (\ref{new14}) and (\ref{new26})
has a unique smooth global solution $\phi(x,t)$
for all $t \in R^1$.}
\end{theorem}

\noindent{\bf Proof:}  The PDE system (\ref{new14})
is manifestly hyperbolic;
hence, local existence and uniqueness are immediate \cite{Sogge}.
To prove global existence, it is sufficient
(by the usual open-closed arguments \cite{Sogge})
to show that if
$\phi(x,t)$ satisfies (\ref{new14}) on $\Sigma \times I$,
with $I$ a bounded open interval in $R^1$,
then $\phi(x,t)$ and all its derivatives are bounded on $\Sigma \times I$.

To show that $\phi$ and its first derivatives $\partial_\alpha\phi$
are bounded,
we use an argument based on stress-energy conservation
(see \cite{Shatah}).
From the form of the
stress-energy components (\ref{new16}) to (\ref{new24}),
together with the conservation equations
\begin{eqnarray}
\partial_t T^t{}_t + \partial_x T^x_t = 0, \qquad
\partial_t T^t{}_x + \partial_xT^x_x =0 , 
\label{new27}
\end{eqnarray}
we find that
\begin{eqnarray}
\gamma^{\alpha \beta} \partial_{\alpha}\partial_{\beta}
 T_{tt} = 0 .
\label{new28}
\end{eqnarray}
It then follows from standard results
(see \cite{Shatah-Struwe}) for the wave equation
on 1 + 1 Minkowski space that $T_{tt}$ is bounded on $I$.
Thus the first derivatives of $\phi$ are bounded.
As a consequence of the mean value theorem
and the assumed compact support of the initial data,
$\phi$ is then bounded as well.

There are a number of ways of proceeding to argue that second
and higher order derivatives of $\phi$ are bounded.
Here we use an argument which is adapted from Shatah \cite{Shatah-Struwe}
based on bounding successive $k$th order energies
\begin{eqnarray}
{\cal E}_k(t) =
{1 \over 2}\int_{\Sigma} \big(
|\partial_t \partial_{x}{}^k\phi|^2
+ |\partial_{x}{}^{k + 1} \phi|^2 \big)dx . 
\label{new29}
\end{eqnarray}
Note that the ordinary energy
\begin{eqnarray}
{\cal E}_0(t) = \int_{\Sigma} T_{tt} dx
= {1 \over 2}\int_{\Sigma} \big(
|\partial_t \phi|^2
+ |\partial_{x} \phi|^2 \big)dx
\end{eqnarray}
is bounded and independent of $t$, 
${\cal E}_0(t) = {\cal E}_0(t_0)$, 
for smooth compact support initial data.

We start by rewriting the torsion wave map equation (\ref{new14})
in the form
\begin{eqnarray}
D^{\alpha} V_\alpha^A = 0
\label{new30}
\end{eqnarray}
where
$V_\alpha^A = \partial_\alpha\phi^A$,
$D^\alpha= \gamma^{\alpha\beta} D_\beta$, 
and
$D_{\beta} = \partial_\beta + \Gamma^A{}_{BC} V_\beta^C
+ \lambda Q^A{}_{BC} \epsilon_\beta{}^\alpha V_\alpha^C$
defines a covariant derivative operator
which includes the connection with torsion.
If we now apply $D_\beta$ to equation (\ref{new30})
and commute $D_{\beta}$ past the derivative operators,
keeping track of the various curvature and torsion terms which arise,
then we obtain a nonlinear wave equation for $V_\alpha^A$:
\begin{eqnarray}
D^{\alpha} D_{\alpha}V_{\beta}{}^A+ P_{\beta}{}^A(V, V,V) = 0
\label{new32}
\end{eqnarray}
where $P(V, V, V)$ denotes an expression which
is trilinear in $V_{\beta}{}^A$
and involves no higher derivatives of $\phi^A$.

By multiplying (\ref{new32}) by
$\gamma^{\alpha\beta} g_{CA} D_t V_{\alpha}^C$,
we straightforwardly derive the conservation equation
\begin{eqnarray}
D_t \big( {1 \over 2} |D_t V|^2 + {1 \over 2} |D_x V| ^2 \big)
- D_x \big( D_t V \cdot D_x V \big)
= \tilde{P}(V, V, V)\cdot DV
\label{new33}
\end{eqnarray}
where $\tilde{P}(V, V, V)$ is, like $P(V, V, V)$,
trilinear in $V$ with no higher derivatives of $\phi^A$.
Now, integrating (\ref{new33}) over $\Sigma$, we obtain
\begin{eqnarray}
\partial_t {\cal E}_1 (t) =
\int_{\Sigma} \tilde{P}(V, V, V)\cdot DV dx
\label{new34}
\end{eqnarray}
for the $1$st order energy defined in (\ref{new29}).
Estimating the right hand side of (\ref{new34}), we find
\begin{eqnarray}
\partial_t {\cal E}_1 (t) \leq C \| V\|_{L^6}^3 \|D_{\alpha} V\|_{L^2}
\nonumber\\
\leq C \| V \|_{L^6}^3 \sqrt{{\cal E}_1 (t)}
\label{new35}
\end{eqnarray}
and hence
\begin{eqnarray}
\partial_t \sqrt{{\cal E}_1} \leq  C \| V \|_{L^6}^3
\quad .
\label{new36}
\end{eqnarray}
It follows from Sobolev inequalities that
\begin{eqnarray}
\| V \|_{L^6} \leq C \| V \|_{L^2}^{2/3} \| DV \|_{L^2}^{1/3} , 
\label{new37}
\end{eqnarray}
so we have
\begin{eqnarray}
\partial_t \sqrt{{\cal E}_1} \leq  C \| V \|_{L^2}^2 \|DV \|_{L^2}
\leq C {\cal E}_0 \sqrt{{\cal E}_1} . 
\label{new38}
\end{eqnarray}
Since ${\cal E}_0$ is bounded, it follows from (\ref{new38}) that
\begin{eqnarray}
 \sqrt{{\cal E}_1(t)} \leq Ce^{kt}
\label{new39}
\end{eqnarray}
which bounds ${\cal E}_1(t)$,
and therefore bounds the $L^2$ norm of $DV$.
Hence $\| \partial_\alpha^2 \phi \|_{L^2}$ is bounded.

To bound ${\cal E}_2(t)$,
we start from the wave equation (\ref{new32}) for $V$
and repeat the previous argument.
Setting $W_{\beta\gamma}{}^A := D_{\beta} V_{\gamma}{}^A$, we derive
\begin{eqnarray}
D^{\alpha}D_{\alpha}W_{\beta \gamma}{}^A
+ {R}_{\beta\gamma}{}^A (V, V, W) = 0
\label{new40}
\end{eqnarray}
where ${R}(V,V,W)$ is bilinear in $V$, linear in $W$, and
involves no other derivatives of $\phi$.
From (\ref{new40}) we obtain the conservation equation
\begin{eqnarray}
D_t \big({1  \over  2}|D_t W |^2 +{1 \over2}|D_x W |^2 \big)
- D_x \big( D_t W \cdot D_xW \big) = \tilde{R}(V, V, W) \cdot DW 
\label{new41}
\end{eqnarray}
where $\tilde{R}(V,V,W)$ has the same properties as $R(V,V,W)$.
Integrating over $\Sigma$ and estimating, we obtain
\begin{eqnarray}
\partial_t {\cal E}_2 (t) \leq C \|V \|_{L^8}^2
\|W \|_{L^4}  \sqrt{ {\cal E}_2(t) }.
\label{new42}
\end{eqnarray}
Since $V$ and $W = DV$ are $L^2$ bounded,
by applying Sobolev inequalities to (\ref{new42})
we obtain
\begin{eqnarray}
\partial_t \sqrt{{\cal E}_2} \leq C \sqrt{{\cal E}_2}
\label{new43}
\end{eqnarray}
Hence we have that ${\cal E}_2(t)$ is bounded
and therefore so is the $L^2$ norm of $DW$.
Thus, since $DW= DDV$, it follows that $\| \partial_\alpha^3\phi\|_{L^2}$
is bounded.

The argument proceeds to all successively higher orders
and we thereby determine that all derivatives of $\phi$ are $L^2$ bounded.
It follows from Sobolev embedding that all derivatives of $\phi$
are pointwise bounded, which completes the proof of Theorem 1. \QED

\section{\head{EQUIVARIANT WAVE MAPS WITH TORSION\\
AND THE FRAME FORMULATION}}

We begin by setting up a frame formulation for wave maps
with and without torsion.
(See also \cite{ChristodoulouT-Z}).
We first choose a frame basis $\{e^A_a\} (a = 1,\dots,n)$
for the target geometry $(N^n,g)$
and let $e^a_A(\psi)$ denote the frame associated to $\psi$. 
We now define the ``frame fields''
\begin{eqnarray}
K^a_{\mu} := e^a_A(\psi) \partial_{\mu}\psi^A
\label{new44}
\end{eqnarray}
where $\{e^a_A\}$ are the components of the dual basis to $\{e^A_a\}$.
These frame fields $K^a_{\mu}$ may be viewed either as
the pull-back of the dual frame $\{ e^a_A \}$ from $N^n$ to $M^{2+1}$
along the map $\psi:M^{2+1} \rightarrow N^n$,
or as the frame components of the wave map gradient 
on the tangent space of 
the target geometry $(N^n,g)$.
In any case, it follows from (\ref{new44}) that
$K$ satisfies the identity
\begin{eqnarray}
\nabla_{[\nu} K^a_{\mu]} = -1/2 C_{bc}{}^a(\psi) K^b_\nu K^c_\mu
\label{Kid}
\end{eqnarray}
where $C_{bc}{}^a$ are the frame commutator coefficients defined by
\begin{eqnarray}
[e_b,e_c]=C_{bc}{}^a e_a .
\label{C}
\end{eqnarray}
Moreover, one verifies that
if $\psi$ satisfies the wave map equation (\ref{wavemapeq}),
then $K$ satisfies
\begin{eqnarray}
\nabla^\mu K^a_\mu = -C^a{}_{bc}(\psi) K^b_\nu
K^c_\mu \eta^{\nu\mu}
\label{divKid}
\end{eqnarray}
where $C^a{}_{bc}:= g^{ad} C_{db}{}^e g_{ec}$,
with $g^{ad}:= e^a_A e^b_B g^{AB}$
and $g_{ab}:= e_a^A e_b^B g_{AB}$;
or if $\psi$ satisfies the torsion wave map equation (\ref{torwavemapeq}),
then $K$ satisfies
\begin{eqnarray}
\nabla^\mu K^a_\mu =
-C^a{}_{bc}(\psi) K^b_\nu K^c_\mu \eta^{\nu\mu}
-\lambda \epsilon^{\sigma\nu\mu}
v_\sigma Q^a{}_{bc}(\psi) K^b_\nu K^c_\mu
\label{tordivKid}
\end{eqnarray}
where $Q^a{}_{bc} := e^a_A Q^A{}_{BC} e^B_b e^C_c$.

Up to this point in setting up the frame formulation,
we have made no restrictions on the choice of the target or
on the nature of the wave maps.
We now focus on equivariant wave maps
(\ref{leftequivar}) to (\ref{conjequivar})
and their corresponding frame formulations,
so we assume the target geometry to be a Lie group $G$.
While $K$ can be defined for any frame basis on $G$, the
frame field equations are simplest if we require that
$\{e^A_a\}$ be a left-invariant basis for $G$.  It then follows that
the commutator coefficients $C_{bc}{}^a$ are independent of $\psi$
and are {\it constant}.
If we make the further restrictions that
the metric $g$ be a left-invariant tensor on $G$,
\begin{eqnarray}
g_{AB} = e^a{}_A e^b{}_B g_{ab} ,
\label{new48}
\end{eqnarray}
and the torsion potential $p$ be a left invariant two-form on $G$,
\begin{eqnarray}
p_{AB} = e^a{}_A e^b{}_B p_{ab},
\label{new49}
\end{eqnarray}
so that the components $g_{ab}$ and $p_{ab}$ are {\it constant},
then the coefficients $C^a{}_{bc}(\psi)$ are 
independent of $\psi$ and {\it constant}
while so are the frame components $Q^a{}_{bc}(\psi)$ as well; 
in particular, we have
\begin{eqnarray}
Q^a{}_{bc} = - 3/2 g^{ad} p_{e[d} C_{bc]}{}^e 
\label{new50}
\end{eqnarray}
with 
\begin{eqnarray}
C_{bc}{}^a = 2e^a_A e^B_{[b|} \partial_B e^A_{|c]} . 
\end{eqnarray}

\noindent{ \bf Remark 1:}
Every nonabelian Lie group admits both
a left-invariant metric $g$ and a left-invariant two-form $p$.
However, for semi-simple Lie groups $G$,
if $G$ has dimension three 
then all left-invariant two-forms $p$ are necessarily closed,
and consequently $Q=0$ so there is no torsion.
This is not the case if $G$ has larger dimension.
In particular, a non-closed left-invariant two-form $p$
and hence non-zero torsion $Q$
is admitted by all nonabelian semi-simple Lie groups $G$
other than the three-dimensional ones
(namely $SU(2)$ and its real forms $SO(3)$, $SO(1,2)$, $SO(2,1)$).
See Proposition~A in the appendix.

We now find that, assuming the restrictions just noted,
we can write equations (\ref{Kid}), (\ref{divKid}) and (\ref{tordivKid})
strictly in terms of the frame fields $K$,
with no explicit $\psi$ dependence:
\begin{eqnarray}
&& \nabla_{[\nu} K^a{}_{\mu]} = -1/2
C_{bc}{}^a K_{\nu}{}^b K^c_{\mu} ,
\label{new51}\\
&& \nabla^{\mu} K^a_{\mu} = -C^a{}_{bc} K^b{}_{\nu} K^c_{\mu}
\eta^{\nu\mu} ,
\label{new52}\\
&& \nabla^{\mu} K_{\mu}{}^a =
-C^a{}_{bc} K^b{}_\nu K^c_{\mu} \eta^{\nu\mu}
-\lambda \epsilon^{\sigma\nu\mu} v_{\sigma} Q^a{}_{bc} K^b{}_\nu K^c_{\mu} .
\label{new54}
\end{eqnarray}
The field equations (\ref{new51}) and(\ref{new52})
together are a self-contained PDE system for $K$
which is equivalent to the wave map equation (\ref{wavemapeq});
the field equations (\ref{new51}) and(\ref{new54})
likewise are a self-contained PDE system for $K$
which is equivalent to the wave map equation 
with torsion (\ref{torwavemapeq}).
Note that the system with torsion reduces to the system without torsion
when $\lambda=0$.

\begin{theorem}
{Proposition 1.}{
Let $(M^{2 + 1},\eta)$ be a Lorentzian geometry,
and let $N^n=G$ be a Lie group target.
\begin{enumerate}
\item  Suppose that $\psi ^A$ is a solution of the torsion wave
map equation (\ref{torwavemapeq}).  Then $K^a_\mu$ defined by
(\ref{new44}) satisfies the field equations (\ref{new51})
and (\ref{new54}). 
\item  Suppose that $K^a_\mu$ is a solution of the field equations
(\ref{new51}) and (\ref{new54}).  If $M^{2 + 1}$ is simply connected,
then there exists a torsion wave map $\psi ^A$,
satisfying equation (\ref{torwavemapeq}), which is related to
$K^a_\mu$ by (\ref{new44}).
\end{enumerate} }
\end{theorem}

\noindent{\bf Proof:}  To prove part (1),
we first note that for $K^a{}_\mu$ given by (\ref{new44}),
the field equation (\ref{new51}) is an identity.
We then verify that, 
through the torsion wave map equation (\ref{torwavemapeq}),
the substitution of (\ref{new44}) for 
$K^a_\mu $ satisfies the field equation (\ref{new54}).

For the converse, to prove part (2), we note
the field equation (\ref{new51}) shows that $K^a_\mu $ can be
viewed as a Lie-algebra valued connection one-form on the
trivial bundle
$M^{2+ 1} \times G$, with zero curvature.  Since the bundle is trivial
and the manifold $M^{2 + 1}$ is assumed to be simply connected,
there exists a global parallel section.  Correspondingly, there exists
a smooth map $U: M^{2 + 1} \rightarrow G$
(called a ``gauge transformation'')
in terms of which we have
\begin{eqnarray}
K^a_\mu  =  (U^{-1} \partial_\mu U  )^A e^a_A (I)
\label{Kgauge}
\end{eqnarray}
where $I$ is the identity element of $G$.

Now let $\psi^A$ denote $U$ written in terms of local
coordinates on $G$.  It follows that $U^{-1}$ pulls back $e^a_A (I)$
to $e^a_A (\psi)$, at the Lie group element specified by $U$.
Hence we have
\begin{eqnarray}
(U^{-1} \partial_\mu U  )^A e^a_A (I) =
e^a_A (\psi) \partial_\mu  \psi^A . 
\label{pullback}
\end{eqnarray}
Combining (\ref{pullback}) with (\ref{Kgauge}),
we obtain equation (\ref{new44}).
Then by substituting (\ref{new44}) into the field equation
(\ref{new54}), we verify that $\psi$ satisfies
(\ref{torwavemapeq}). \QED

We note that, independent of their usefulness for the study of wave maps, 
these field theories 
in terms of $K$ viewed as a Lie-algebra valued one-form field on $M^{2+1}$
have some interest as a nonlinear
generalization of Maxwell's equations. Indeed, for the abelian case
$C_{ab}{}^c=0$, the field equations (\ref{new51}) and
(\ref{new52}) are exactly Maxwell's equations in 2+1
dimensions, while the field equations (\ref{new51}) and
(\ref{new54}) are a modification of  Maxwell's equations by
adding torsion. This relationship is explored elsewhere
\cite{Anco,Anco2,AncoIsenberg}.

Since we will use frame fields to study translation equivariant
wave maps,  we now characterize frame fields which
correspond to the three classes of equivariant wave maps
(\ref{leftequivar}), (\ref{rightequivar}), (\ref{conjequivar}).
We begin with the following definitions of
invariant and equivariant frame fields under a translation group action.

\noindent{\bf Invariant Frame Field:}
\begin{eqnarray}
K (x, y + \lambda, t ) = K (x, y, t )
\label{new57}
\end{eqnarray}
\noindent{\bf Equivariant Frame Field:}
\begin{eqnarray}
K (x, y + \lambda, t ) = \exp(-\lambda A) K (x, y, t) \exp(\lambda A)
\label{new58}
\end{eqnarray}
Here $A$ is an element of the Lie algebra of the target Lie group $G$,
and $(x,y,t)$ are standard coordinates for the Minkowski space base geometry
$(M^{2+1},\eta)$.

Geometrically, the translation equivariant group action (\ref{new58})
on $K$ arises via the pull-back of the dual frame components $\{ e^a_A \}$
under right multiplication in $G$
by the one-parameter exponential subgroup generated from the
Lie algebra element $R$.
When $R=0$ this group action reduces to
the translation invariant group action (\ref{new57}) on $K$.
(Alternatively, note that
the translation invariant group action arises directly
by left multiplication in $G$ since the dual frame is left-invariant.)

Based on Proposition~1,
the correspondence between invariant/equivariant frame fields and wave maps
is summarized by the following two results.

\begin{theorem}
{Proposition 2.}{
\begin{enumerate}
\item If $\psi$ is left equivariant (\ref{leftequivar}),
then the corresponding frame field $K$ is invariant (\ref{new57}).
\item If $\psi$ is right equivariant (\ref{rightequivar}),
then the corresponding frame field $K$ is equivariant (\ref{new58}),
with the components $K^a_y$ constant.
\item If $\psi$ is conjugate equivariant (\ref{conjequivar}),
then the corresponding frame field $K$ is equivariant (\ref{new58}).
\end{enumerate} }
\end{theorem}

The proof of these correspondences amounts to a direct calculation
using a matrix representation for $\psi$ and $K$.
There are straightforward converse correspondences as well. 

\begin{theorem}
{Proposition 3.}{
\begin{enumerate}
\item If K is invariant (\ref{new57}), then the corresponding wave map $\psi$
is left equivariant (\ref{leftequivar}).
\item If K is equivariant (\ref{new58}), then the corresponding wave map $\psi$
is conjugate equivariant (\ref{conjequivar}).
\item If K is equivariant (\ref{new58}) with the components $K^a_y$ constant,
then the corresponding wave map $\psi$ 
is right equivariant (\ref{rightequivar}).
\end{enumerate}}
\end{theorem}

\noindent{\bf Proof:}
Let $U$ denote a the matrix representation of the
wave map $\psi$ corresponding to $K$.
We first prove part (1).
It follows from the definition of frame field invariance,
together with relation (\ref{Kgauge}),
that $U$ satisfies
\begin{eqnarray}
\partial_y (U^{-1}\partial_\mu U) = 0 . 
\label{new59}
\end{eqnarray}
Integrating the $y$ component of this equation,
and then multiplying both sides by $U$,
we obtain the linear matrix ordinary differential equation
\begin{eqnarray}
\partial_y U (x, y, t) = U (x, y, t) f (x, t)
\label{new60}
\end{eqnarray}
where $f$ is an arbitrary Lie-algebra matrix valued function
(independent of $y$).  The general solution to (\ref{new60}) is
\begin{eqnarray}
 U (x, y, t) = \exp (yA (x,t)) V (x, t)
\label{new61}
\end{eqnarray}
where $V$ is an arbitrary Lie-algebra nonsingular matrix valued function
(independent of $y$),
and $A := V f  V^{-1}$.

We now impose the $x,t$ components of equation (\ref{new59}).
Calculating $U^{-1}\partial_t U$ with $U$ from (\ref{new61}), we find
\begin{eqnarray}
U^{-1}\partial_t U =
V^{-1} \partial_t V + V^{-1} \exp(-yA) ( \partial_t \exp(yA ) ) V ,
\label{new62}
\end{eqnarray}
so
\begin{eqnarray}
0 &=& \partial_y (U^{-1}\partial_t  U)
 = V^{-1}  \exp(-yA) \partial_t A \, \exp(yA) V ,
\label{new63}
\end{eqnarray}
which implies that
\begin{eqnarray}
\partial_t A = 0 .
\label{new64}
\end{eqnarray}
Similarly, working with $U^{-1} \partial_x U$ and imposing $0 =
\partial_y (U^{-1} \partial_x U)$ we determine that
\begin{eqnarray}
\partial_x A= 0 . 
\label{new65}
\end{eqnarray}
Thus $A$ must be a {\it constant} Lie-algebra valued matrix,
which we denote $L$; 
then (\ref{new61}) becomes
\begin{eqnarray}
U(x, y, t) = \exp(yL) V (x, t) .
\label{new66}
\end{eqnarray}
Condition (\ref{leftequivar}) immediately follows, so the wave map
corresponding to $K$ is left equivariant.

We now prove part (2).  From the definition of frame equivariance,
there is a $y$-independent Lie-algebra matrix valued field
$f_{\mu} (x, t)$ and a constant Lie-algebra matrix $A$ such that
\begin{eqnarray}
K_{\mu} (x, y, t) = \exp (-y A) f_{\mu} (x, t) \exp (y A) .
\label{new67}
\end{eqnarray}
Hence, from relation (\ref{Kgauge}),
$U (x, y, t)$ must satisfy
\begin{eqnarray}
U^{-1} \partial_{\mu} U = \exp(-y A) f_{\mu} \exp (y A) .
\label{new68}
\end{eqnarray}
The $y$-component of this equation yields
\begin{eqnarray}
(\partial_y U) \exp (-y A) = U \exp (-y A) f_y , 
\label{new69}
\end{eqnarray}
and after some manipulation we obtain the linear matrix ODE
\begin{eqnarray}
\partial_y W = W (f_y - A)
\label{new70}
\end{eqnarray}
where $W (x, y, t) :=  U (x, y, t) \exp (-y A)$.
The general solution to (\ref{new70}) is
\begin{eqnarray}
W (x, y, t) = \exp (y B(x, t)) V (x, t)
\label{new71}
\end{eqnarray}
where $V $ is an arbitrary Lie-algebra matrix valued function,
and $B$ is defined as
\begin{eqnarray}
B := V(f_y - A)V^{-1} . 
\label{new72}
\end{eqnarray}
Working with the other components of equation (\ref{new68}) we derive
\begin{eqnarray}
\partial_t W = W f_t
\label{new73}
\end{eqnarray}
and
\begin{eqnarray}
\partial_x W = W f_x . 
\label{new74}
\end{eqnarray}
Then rearranging (\ref{new73}) and using (\ref{new71}), we obtain
\begin{eqnarray}
f_t (x, t) &=& W^{-1} \partial _t W
 = V^{-1} \partial_t V + V^{-1}\exp(-y B) \partial _t (\exp(-y B)) V . 
\label{new75}
\end{eqnarray}
Since both $f_t$ and $V$
are independent of $y$, if we take $\partial_y$ of both sides of
equation (\ref{new75}) we have
\begin{eqnarray}
0 &=& \partial_y (V^{-1}\, \exp(-y B) \partial _t (\exp (y B)) V)
\nonumber\\
&=& V^{-1}\exp(-y B) \partial _t B \exp(y B) V
\label{new76}
\end{eqnarray}
which implies that $\partial _t B = 0$.
Similarly, using (\ref{new74}), we find that $\partial _x B = 0$.
Hence $B$ is a {\it constant} Lie-algebra matrix, 
which we denote $L$. 
Thus, after combining (\ref{new71})
with the definition $W=U\exp(-yA)$, we see that
\begin{eqnarray}
U (x, y, t)  = \exp(yL) V (x, t) \exp (y A)
\label{new77}
\end{eqnarray}
so $U (x, y, t)$ is conjugate equivariant (\ref{conjequivar})
with $R=A$. 

Finally, we prove part (3). 
From (\ref{new77}) we have
\begin{eqnarray}
K_y= U^{-1} \partial_y U = \exp(-yA) V^{-1} L V \exp (y A) +A 
\label{new78}
\end{eqnarray}
which is assumed to be constant. 
By differentiating with respect to $y$, we obtain
$[V^{-1} L V , A]=0$, 
and hence (\ref{new77}) becomes 
$K_y= V^{-1} L V +A$. 
Thus, it follows that $B:=V^{-1} L V$ defines 
a constant Lie-algebra matrix which commutes with $A$. 
We then have
\begin{eqnarray}
U (x,y,t) &=& \exp(yL) V(x,t) \exp (y A) 
\nonumber\\
&=& V(x,t) \exp(y B) \exp (y A) 
= V(x,t) \exp(y R)
\label{new79}
\end{eqnarray}
where $R:=A+B$. 
Hence $U(x,y,t)$ is right equivariant (\ref{rightequivar}).\QED

As a consequence of Propositions~2 and~3,
we can prove global existence of solutions to the Cauchy problem
for the three classes of translation equivariant wave maps
(with or without torsion) by using invariant or equivariant frame fields.
We do this first for the invariant frame fields in the next section.

Our analysis makes essential use of
the wave map stress-energy tensor (\ref{T}).
Through the relation (\ref{new44}) for $K$ in terms of $\psi$,
we obtain
\begin{eqnarray}
T^\mu{}_\alpha =
\eta^{\mu\nu} K^a_\nu K^b_\alpha g_{ab}
- 1/2 \delta^\mu{}_\alpha \eta^{\nu\sigma} K^a_\nu K^b_\sigma g_{ab}
+ 1/2 \lambda \epsilon^{\mu\nu\sigma} v_\alpha p_{ab} K^a_\nu K^b_\sigma . 
\label{frameT}
\end{eqnarray}
One verifies that, for solutions $K$ of (\ref{new51}) and (\ref{new54})
in which $(M^{2+1},\eta)$ is Minkowski space,
this non-symmetric stress-energy tensor satisfies the conservation equation
\begin{eqnarray}
\partial_\mu T^\mu{}_\alpha =
1/2 \lambda \epsilon^{\mu\nu\sigma} p_{ab} K^a_\mu K^b_\nu
\partial_\alpha v_\sigma . 
\label{framedivT}
\end{eqnarray}
Hereafter we specialize to the situation where $v$ is constant
on $M^{2+1}$.
This makes the analysis of the field equations
considerably simpler. In particular, the stress-energy is strictly conserved,
$\partial_\mu T^\mu{}_\alpha = 0$.

\section{\head{GLOBAL EXISTENCE FOR 
INVARIANT FRAME FIELD EQUATIONS WITH TORSION}}

By definition (\ref{new57}) of translation invariance for frame fields,
the component functions $K^a_\mu$ are independent of $y$.
Then, adopting the convenient notation
\begin{eqnarray}
E^a := K^a_x,
H^a := K^a_y,
B^a := K^a_t,
\end{eqnarray}
we find that the translation-invariant frame field equations
take the form
\begin{eqnarray}
\partial _x H^a = &&
- C_{bc}{}^a E^b H^c
\label{reducedconstraint}\\
\partial _t E^a = &&
\partial _x B - C_{bc}{}^a B^b E^c
\label{reducedExevolution}\\
\partial _t H^a = &&
- C_{bc}{}^a B^b H^c
\label{reducedEyevolution}\\
\partial _t B^a = &&
\partial _x E^a - C^a{}_{bc} \Big( B^b B^c - E^b E^c - H^b H^c \Big)
\nonumber \\&&\quad
- \lambda Q ^a{}_{bc}
\Big(v_y B^b E^c - v_x B^b H^c + v_t E^b H^c \Big)
\label{reducedBevolution}
\end{eqnarray}
for the functions $\{E^a (x, t), H^a (x, t), B^a (x, t)\}$.
Note that, in this system of field equations,
(\ref{reducedconstraint}) is a constraint equation
while (\ref{reducedExevolution}) to (\ref{reducedBevolution})
are evolution equations.

Initial data at $t=t_0$ for the Cauchy problem is specified by
choosing (on $\Sigma=R^1$ or $S^1$ allowing for periodic boundary conditions) 
Lie-algebra valued functions
$\{ {\hat E}^a(x), {\hat H}^a(x), {\hat B}^a(x) \}$
which satisfy the constraint
\begin{eqnarray}
\partial_x {\hat H}^a = -C_{bc}{}^a {\hat E}^b {\hat H}^c . 
\label{dataconstraint}
\end{eqnarray}
 A solution to the Cauchy problem is then a set of fields 
$\{E^a (x, t)$, $H^a (x, t)$, $B^a (x, t)\}$  
satisfying (\ref{reducedExevolution}) to
(\ref{reducedBevolution})  and the initial conditions
\begin{eqnarray}
E^a(x,t_0) = {\hat E}^a(x),
H^a(x,t_0) = {\hat H}^a(x),
B^a(x,t_0) = {\hat B}^a(x) . 
\label{initialvalues}
\end{eqnarray}

To show that the Cauchy problem is well-posed,
we note that well-posed\-ness is known for
the wave map equation without torsion \cite{Shatah-Struwe},
which is equivalent to the system
(\ref{reducedconstraint}) to (\ref{reducedBevolution})
up to the addition of the torsion terms involving $\lambda$.
These terms do not involve any derivatives of the fields
and hence do not effect the well-posedness.
Alternatively, we note that, up to such terms, the system is
equivalent to the Maxwell equations in 2+1 dimensions,
which constitute a well-posed system.
It follows that the system
(\ref{reducedconstraint}) to (\ref{reducedBevolution})
is well-posed and, moreover, is first-order hyperbolic.

In this section we prove global existence of smooth solutions
to the Cauchy problem for the 1+1 field equations
(\ref{reducedconstraint}) to (\ref{reducedBevolution}).
The proof relies on the use of the stress-energy tensor (\ref{frameT})
along with light cone estimates.

To proceed we write out the
components of the stress-energy tensor (\ref{T})
in terms of $E^a, H^a, B^a$.
Using the coordinates $(x, y, t)$ for $M^{2+1}$ we have
\begin{eqnarray}
&& T_{xx} = {1 \over 2} \big( E^2_x - E^2_y + B^2 \big)
+ \lambda v_x H^a B^b p_{ab}
\label{Txx}\\
&& T_{yy} = {1 \over 2} \big(-E^2_x + E^2_y + B^2 \big)
+\lambda v_y B^a E^bp_{ab}
\label{Tyy}\\
&& T_{tx} = E\cdot B + \lambda v_x H^a E^b p_{ab}
\label{Ttx}\\
&& T_{xt} = E\cdot B + \lambda v_t H^a B^b p_{ab}
\label{Txt}\\
&& T_{ty} = H\cdot B +  \lambda v_y H^a E^b p_{ab}
\label{Tty}\\
&& T_{yt} = H\cdot B +  \lambda v_t B^a E^b p_{ab}
\label{Tyt}\\
&& T_{xy} = E\cdot H +  \lambda v_y H^a B^b p_{ab}
\label{Txy}\\
&& T_{yx} = E\cdot H + \lambda v_x B^a E^b p_{ab}
\label{Tyx}
\end{eqnarray}
where $E^2: = E^a E^b g_{ab}$ and $E\cdot B = E^a B^b g_{ab}$, etc..

For derivation of light cone estimates,
it is useful to work with null components of the stress-energy tensor.
We introduce null coordinates which mix $t$ and $x$ (but not $y$):
\begin{eqnarray}
\begin{array}{ccc}
\ell = t + x&&x = {1 \over 2} (\ell + n)\\
&\longleftrightarrow&\\
n = -t +x&&t = {1 \over 2} (\ell - n)\\
\end{array} \quad .
\end{eqnarray}
Then we find (for the components we will need):
\begin{eqnarray}
T_{\ell \ell} &=& K^2_{\ell} + \lambda v_{\ell}H^a K^b_{\ell} p_{ab}
\nonumber \\  &=&
{1\over 4}(B + E)^2 + {\lambda \over 4} (v_t + v_x)H^a
(B^b + E^b )p_{ab}
\label{Tll}
\end{eqnarray}
\begin{eqnarray}
T_{nn} &=& K^2_n - \lambda v_n H^a K^b_n p_{ab}
\nonumber \\  &=&
{1\over 4}(-B + E)^2 - {\lambda \over 4} (-v_t + v_x) H^a
(-B^b + E^b ) p_{ab}
\label{Tnn}
\end{eqnarray}
\begin{eqnarray}
T_{\ell n} &=&-{1 \over 2} K^2_y - \lambda v_n K^a_{\ell}
H^b p_{ab} \nonumber \\  &=&
-{1\over 2} E^2_y + {\lambda \over 4} (-v_t + v_x) H^a
(B^b +E^b )p_{ab}
\label{Tln}
\end{eqnarray}
\begin{eqnarray}
T_{n \ell} &=& -{1 \over 2} K^2_y + \lambda v_{\ell} K^a_n
H^b p_{ab} \nonumber \\  &=&
-{1\over 2} E^2_y - {\lambda \over 4} (v_t + v_x) H^a
(-B^b +E^b )p_{ab} .
\label{Tnl}
\end{eqnarray}
For these components the stress-energy conservation equation
(\ref{framedivT}) has the null component form
\begin{eqnarray}
\partial_n T_{\ell \ell} + \partial_{\ell}
T_{n \ell} = 0 , 
\label{new95}
\end{eqnarray}
\begin{eqnarray}
\partial_{\ell} T_{nn} + \partial_{n} T_{\ell n} = 0 .
\label{new96}
\end{eqnarray}
These equations are essential for the derivation of the light
cone estimates we will need.

Also important for our analysis is the
energy function
\begin{eqnarray}
{\cal E}(t) &=& \int_\Sigma T_{tt} dx\nonumber\\
&=& \int_\Sigma \big(
{1 \over 2} (E{}^2 + H{}^2 + B^2) + \lambda v_t H^a E^b p_{ab} \big)dx
\label{new97}
\end{eqnarray}
We note that for certain values of the coupling constant $\lambda$,
the energy ${\cal E} (t)$ can be negative,
and it therefore does not in general control the
$L^2$ norm of $E^a$, $H^a$, or $B^a$.
However, for sufficiently small $\lambda$,
there is a constant $k > 0$ such that
\begin{eqnarray}
{1  \over  k}  (E^2_x + E^2_y  ) \leq E^2_x + E^2_y
+ 2\lambda v _t H^a E^b p_{ab} \leq k  (E^2_x + E^2_y )
\label{Ebound}
\end{eqnarray}
and hence the energy is positive, so that ${\cal E} (t)$
{\it does} consequently control
$\|E\|_{L^2}$,$\|H\|_{L^2}$ , and $\|B\|_{L^2}$.
We assume henceforth that $\lambda$ is sufficiently small for
this to be the case.
\footnote{It is sufficient that $\lambda$ satisfy
\begin{eqnarray*}
|\lambda| \leq  1/\sqrt{ |v_t|  |p| }
\end{eqnarray*}
where $|p|^2 = |p_{ab}p_{cd} g^{ac} g^{bd}|$. }

We now state our main results.
Let $\Sigma$ denote $R^1$ or $S^1$,
and introduce coordinates $(x,t)$ for $\Sigma\times R^1 \simeq M^{2+1}$.
Fix constants $v_t,v_x,v_y$.
Let $G$ be a Lie group with $C_{bc}{}^a$ denoting the
Lie-algebra commutator structure tensor.
Fix on the Lie algebra of $G$ a positive definite metric tensor
$g_{ab}$ (it need not necessarily be compatible with the commutator)
and a skew-tensor $p_{ab}$.
Let $Q^a{}_{bc}$ be the tensor defined by (\ref{new50}).

\begin{theorem}
{Theorem 2.}{
Let $\lambda$ be a small constant.${}^4$
For any smooth compact support initial data (\ref{initialvalues})
satisfying (\ref{dataconstraint}),
the Cauchy problem (\ref{reducedconstraint}) to (\ref{reducedBevolution})
has a unique smooth global solution
$\{E^a (x, t), H^a (x, t),  B^a (x, t)\}$ for all $t \in R^1$ }
\end{theorem}

Combining this result with Propositions~2 and~3 from Section~3,
we have a corresponding result for wave maps.

\begin{theorem}
{Theorem 3.}{
The Cauchy problem for left-translation equivariant
Lie group wave maps (\ref{leftequivar}), with or without torsion,
has a unique smooth global solution for all smooth compact support
initial data. }
\end{theorem}

\noindent{\bf Proof of Theorem 2:}

Local existence and uniqueness of smooth solutions of
the PDE system (\ref{reducedconstraint}) to (\ref{reducedBevolution})
follows from standard results (see, for example,\cite{Sogge})
for first-order hyperbolic systems in 1+1 dimensions.
In order to prove global existence,
it is sufficient by the usual ``open-closed'' arguments \cite{Sogge}
to establish the following:
For $\{ E^a (x, t)$, $H^a (x, t)$, $B^a (x, t) \}$
satisfying equations (\ref{reducedconstraint}) to
(\ref{reducedBevolution}) for $t \in I$, with $I$ a bounded open
interval in $R^1$, each component of these fields is {\it bounded}
for $t \in I$, as are all orders of their derivatives.
We prove this boundedness result as follows:

\subsubsection*{Step 1: Conserved Energy}

It follows from the stress-energy conservation equation
$\partial_t T_{tt} - \partial_x T_{xt}=0$ that
the energy  ${\cal E}(t)$ satisfies
${d \over dt} {\cal E} (t) = {\cal F} (t)$
where
\begin{eqnarray}
{\cal F} (t) := \int_\Sigma \partial_x T_{xt} dx
= T_{xt} \Big|_{\partial\Sigma}
\label{new101}
\end{eqnarray}
is the flux.
If we are working on $\Sigma=S^1$,
then $\partial\Sigma$ is empty, so ${\cal F} (t)=0$.
If instead $\Sigma=R^1$,
then we note that as a consequence of hyperbolicity of the system
(\ref{reducedconstraint}) to (\ref{reducedBevolution}),
the fields $\{ E^a, H^a, B^a \}$ have compact support on $\Sigma$
for all $t\in I$,
and hence ${\cal F} (t)=0$.
Thus, the energy is conserved,
${\cal E} (t) = {\cal E} (t_0)$,
for all $t\in I$.

As we noted earlier,
the energy controls the $L^2$ norm of the fields $\{ E^a$, $H^a$, $B^a \}$,
so long as $\lambda$ is sufficiently small (as assumed in the theorem).
Hence we have
\begin{eqnarray}
\| E^a \|_{L^2 (\Sigma  )}<k, \quad
\| H^a \|_{L^2 (\Sigma  )}<k, \quad
\| B^a \|_{L^2 (\Sigma  )}<k
\label{Ltwo}
\end{eqnarray}
for some constant $k$ (depending on ${\cal E} (t_0)$),
for all $t \in I$.

\subsubsection*{Step 2: Bounded $H^a$}

In the system (\ref{reducedconstraint}) to (\ref{reducedBevolution}),
the field $H^a$ enters in a different way from $E^a$ and $B^a$,
since the evolution equation (\ref{reducedEyevolution}) for $H^a$
involves no spatial derivative terms,
and the constraint equation (\ref{reducedconstraint}) has
$\partial_x H^a$ appearing, but no spatial derivatives of
$E^a$ or $B^a$.
Consequently, we treat $H^a$ differently from the other two fields:
we first show that $H^a$ is bounded, and then use this in
showing that $E^a$ and $B^a$ are bounded.

To start, we integrate the absolute values of both sides of
the constraint equation (\ref{reducedconstraint}) over $\Sigma$,
obtaining
\begin{eqnarray}
\int_{\Sigma}  |\partial_x H^a  |dx =
\int_{\Sigma} | C_{bc}{}^a E^b H^c | dx .
\label{LonexderEy}
\end{eqnarray}
Since $C_{bc}{}^a$ is constant, there exists a constant $k_1$
such that
\begin{eqnarray}
| C_{bc}{}^a E^b H^c | \leq
| C_{bc}{}^a | |E^b| |H^c| \leq
 k_1 (E^2 + H^2)
\label{ExEyest}
\end{eqnarray}
by standard algebraic inequalities.
It follows from (\ref{LonexderEy}) and (\ref{ExEyest})
together with the bounds (\ref{Ltwo}) that
\begin{eqnarray}
\int_{\Sigma}  |\partial_x  H^a | dx \leq k_2
\label{LoneEyest}
\end{eqnarray}
for a constant $k_2$.
Combining (\ref{LoneEyest}) with the mean value theorem,
we obtain controls on the spatial variation of $E_y(x,t)$
for any fixed time $t$.  In particular,
for any $x_1,x_2 \in \Sigma$ with fixed $t$,
we have
\begin{eqnarray}
 | H^a  (x_2, t  ) - H^a  (x_1, t  ) | &=&
 | \int^{x_2}_{x_1}\partial_x  H^a(x,t) dx  | \nonumber \\
&\leq&
\int^{x_2}_{x_1}  |\partial_x  H^a(x,t) | dx \leq k_2 .
\label{varEyest}
\end{eqnarray}

If we are working on $\Sigma = R^1$,
we can choose $x_1$ outside the support of $H^a(x,t)$ for all $t \in I$,
and therefore it follows from (\ref{varEyest}) that
$ | H^a(x,t) | \leq k_2 $ for all $(x,t) \in \Sigma \times I$.
Hence, $H^a(x,t)$ is bounded on $\Sigma \times I$.

If instead we are working on $\Sigma = S^1$, we need to do more
to bound $H^a(x,t)$.
Consider $\int_{S^1}H^a(x,t) dx$,
which is the spatial average of $H^a$ on $S^1$.
From the fundamental theorem of calculus,
and from the evolution equation (\ref{reducedEyevolution}),
we obtain (for $t \in I$)
\begin{eqnarray}
\int_{S^1}H^a(x,t)dx &=& \int^t_{t_0} {d  \over  d s}
\int_{S^1}H^a(x,s)dx ds + \int_{S^1}H^a(x,t_0)dx
\nonumber \\
&=& - \int^t_{t_0}\int_{S^1} C_{bc}{}^a B^b(x,s) H^c(x,s) dx ds +
\int_{S^1}H^a (x,t_0 )dx .
\nonumber\\
\label{LoneSEy}
\end{eqnarray}
Next, using standard quadratic algebraic inequalities,
we note that \break $\int_{S^1} C_{bc}{}^a B^b H^c dx$ is bounded
in terms of the energy,
\begin{eqnarray}
| \int_{S^1}C^a_{bc} B^b H^cdx |
\leq k_3 {\cal E} (t) = k_3 {\cal E} (t_0 )
\label{LoneS1BEy}
\end{eqnarray}
for some constant $k_3$.
Hence, $\int^t_{t_0}\int_{S^1} C_{bc}{}^a B^b(x,s) H^c(x,s) dx ds$
is bounded above and below,
\begin{eqnarray}
| \int^t_{t_0}\int_{S^1} C_{bc}{}^a B^b(x,s) H^c(x,s) dx ds |
\leq (t-t_0) k_3 {\cal E} (t_0 ) \leq k_4
\label{new109}
\end{eqnarray}
for some constant $k_4$,
for all $t\in I$.
Then since $\int_{S^1}H^a (x,t_0 )dx$ involves initial data only,
it also is bounded above and below.
Therefore, from (\ref{LoneSEy}) we have that
\begin{eqnarray}
| \int_{S^1}H^a (x,t )dx | \leq k_5
\label{new110}
\end{eqnarray}
and so the average of $H^a$ over $S^1$ is bounded above and below,
for all $t \in I$.  Combining this result with the spatial variance control
(\ref{varEyest}), we conclude that $H^a(x,t)$ is bounded (above and below)
on $\Sigma \times I$.

\subsubsection*{Step 3: Bounded $E^a$ and $B^a$}

While standard $1 + 1$ light cone arguments do not directly
apply to the system (\ref{reducedconstraint}) to (\ref{reducedBevolution}),
a modified argument can be used with the pointwise bounds
on $H^a$ achieved in Step 2.

Using the null form of the stress-energy conservation laws
(\ref{new95})-(\ref{new96}), along with the expressions
(\ref{Tll})-(\ref{Tnl}) for the stress-energy components,
we have
\begin{eqnarray}
\partial_n  (B + E  )^2 &=&
2 \partial_{\ell} H^2
- \lambda  (v_t+ v_x ) p_{ab}
\partial_n \Big( H^a  (B^b + E^b ) \Big ) \nonumber \\
&& + \lambda  (v_t + v_x ) p_{ab}
\partial_{\ell} \Big( H^a  (-B^b +E^b ) \Big) ,
\label{new111}
\end{eqnarray}
\begin{eqnarray}
\partial_{\ell}  (-B + E  )^2  &=&
2 \partial_n H^2
- \lambda  (-v_t + v_x ) p_{ab}
\partial_{n}  \Big( H^a  (B^b +E^b ) \Big) \nonumber \\
&& + \lambda  (-v_t+ v_x ) p_{ab}
\partial_{\ell}  \Big( H^a  (-B^b +E^b ) \Big) .
\label{new112}
\end{eqnarray}
We use the field equations
(\ref{reducedconstraint}) to (\ref{reducedBevolution})
to remove all of the derivatives which appear on the right-side of
these equations.  Thus
\begin{eqnarray}
\partial_n  (B + E  )^2 &=&
-2H^b H^c  (B^a + E^a ) C_{abc}
\nonumber \\ &&
-2 \lambda  (v_t +v_x ) H^a  (-B^b + E^b )  (B^c +E^c )Q_{bca}
\label{new113}
\end{eqnarray}
and
\begin{eqnarray}
\partial_\ell  (-B + E  )^2 &=&
-2 H^b H^c  (-B^a + E^a ) C_{abc}
\nonumber \\  &&
-2 \lambda  (-v_t +v_x ) H^a  (-B^b + E^b )  (B^c +E^c )Q_{bca} .
\label{new114}
\end{eqnarray}
It is convenient here to let
$\alpha^a := B^a + E^a$ and $\beta^a := - B^a + E^a$,
and so we have
\begin{eqnarray}
\partial_n \alpha^2 = -2C_{abc} H^b H^c \alpha^a
-2 \lambda (v_t + v_x )Q_{bca} H^a \beta^b \alpha^c
\label{nderalphaeq}
\end{eqnarray}
and
\begin{eqnarray}
\partial_{\ell} \beta^2 = -2 C_{abc} H^b H^c \beta^a
-2\lambda (-v_t + v_x )Q_{bca} H^a \beta^b \alpha^c .
\label{lderbetaeq}
\end{eqnarray}
Since $C_{abc}$, $Q_{bca}$, $\lambda$, $v_t$ and $v_x$ are constant,
and since $H^a$ is bounded on $\Sigma \times I$,
we immediately have the following estimates
for the right-sides of (\ref{nderalphaeq}) and (\ref{lderbetaeq}):
\begin{eqnarray}
\partial_n \alpha^2 \leq
k_6 \sqrt{\alpha^2} + k_7 \sqrt{\alpha^2}\sqrt{\beta^2}
\label{nderalpha}
\end{eqnarray}
and
\begin{eqnarray}
\partial_{\ell} \beta^2 \leq
k_8 \sqrt{\beta^2} + k_9 \sqrt{\alpha^2}\sqrt{\beta^2}
\label{lderbeta}
\end{eqnarray}
with some constants $k_6$, $k_7$, $k_8$, and $k_9$.

We now apply a light cone argument to the differential inequalities
(\ref{nderalpha}) and (\ref{lderbeta}).
First, choose an arbitrary point $(\hat{x}, \hat{t})$ in $\Sigma\times I$
to the future of the initial surface $\Sigma$, so $\hat t > t_0$,
and integrate $\alpha^2$ back along the light ray
parallel to $\partial_n$ via (\ref{nderalpha})
and also integrate $\beta^2$ back along the light ray
parallel to $\partial_{\ell}$ via (\ref{lderbeta}).
This yields
\begin{eqnarray}
\alpha^2 (\hat{x},  \hat{t} ) \leq \alpha^2  (\hat{x} +
\hat{t} - t_0, t_0 ) + k_6 \int^{ \hat{t}}_{t_0}
\sqrt{\alpha^2 (\hat{x} + \hat{t} - s,s )} ds \nonumber \\
+ k_7 \int^{ \hat{t}}_{t_0}
\sqrt{\alpha^2 (\hat{x} + \hat{t} - s,s )}\,
\sqrt{\beta^2 (\hat{x} + \hat{t} - s,s )}ds
\label{alphaeq}
\end{eqnarray}
and
\begin{eqnarray}
\beta^2 (\hat{x},  \hat{t} ) \leq \beta^2  (\hat{x} +
 t_0 - \hat{t}, t_0 ) + k_8 \int^{ \hat{t}}_{t_0}
\sqrt{\beta^2 (\hat{x} -\hat{t} + s,s )} ds \nonumber \\
+ k_9 \int^{ \hat{t}}_{t_0}
\sqrt{\alpha^2 (\hat{x} - \hat{t} + s,s )} \,
\sqrt{\beta^2 (\hat{x} - \hat{t} + s,s )}ds .
\label{betaeq}
\end{eqnarray}
Next, take the supremum of these expressions over $\Sigma$.
Letting
$\hat{\alpha}^2 (t) := \sup_{x\in \Sigma} \alpha^2(x,t)$
and $\hat{\beta}^2 (t) := \sup_{x\in \Sigma} \beta^2(x,t)$,
we obtain from (\ref{alphaeq})
\begin{eqnarray}
\hat{\alpha}^2  ( \hat{t} ) &\leq&
{\hat\alpha}^2  ( t_0 ) + k_6 \sup_{x\in \Sigma}\int^{ \hat{t}}_{t_0}
\sqrt{\alpha^2 (x ,s )} ds \nonumber \\
&&+ k_7 \sup_{x\in \Sigma} \int^{ \hat{t}}_{t_0}
\sqrt{\alpha^2 (x ,s )}\,
\sqrt{\beta^2 (x ,s )}ds  \nonumber \\
&\leq& {\hat\alpha}^2 (t_0 )
+k_6 \int^{ \hat{t}}_{t_0}\, \sqrt{\hat{\alpha}^2 (s )} ds
+ k_7  \int^{ \hat{t}}_{t_0}
\sqrt{\hat{\alpha}^2 (s )}\, \sqrt{\hat{\beta}^2 (s )}ds \nonumber \\
&\leq& {\hat\alpha}^2 (t_0 )
+k_{10}  ( \hat{t}-{t_0} )^{1/2} \Big( \int^{\hat{t}}_{t_0}
\hat{\alpha}^2  (s ) ds \Big)^{1/2} \nonumber \\
&&+ k_{11}  \Big( \int^{ \hat{t}}_{t_0}
\hat{\alpha}^2 (s )ds \Big)^{1/2}\,
\Big( \int^{\hat{t}}_{t_0} \hat{\beta}^2 (s )ds \Big)^{1/2}
\label{hatalphaeq}
\end{eqnarray}
where the last step is a consequence of the Holder inequality.
If we define
\begin{eqnarray}
a(t) := \int^{t}_{t_0} \hat{\alpha}^2 (s )ds
\label{a}
\end{eqnarray}
and
\begin{eqnarray}
b(t) := \int^{t}_{t_0} \hat{\beta}^2 (s )ds
\label{b}
\end{eqnarray}
then (\ref{hatalphaeq}) can be written as
(with $\hat{t}$ replaced by $t$)
\begin{eqnarray}
{d \over  dt} \, a(t) \leq a(t_0) + k_{10}  (t - t_0 )^{1/2}a^{1/2}(t)
+ k_{11}a^{1/2}(t)b^{1/2}(t) .
\label{aeq}
\end{eqnarray}
Similarly, from (\ref{betaeq}), we derive
\begin{eqnarray}
{d \over  dt} \, b(t) \leq b(t_0) + k_{12}  (t - t_0 )^{1/2}b^{1/2}(t)
+ k_{13}a^{1/2}(t)b^{1/2}(t) .
\label{beq}
\end{eqnarray}

We want to show $a(t)$ and $b(t)$ are bounded functions of $t$
by applying a Gronwall type argument to the coupled inequalities
(\ref{aeq}),(\ref{beq}).  It is useful first to divide
by $a^{1/2}(t)$ in (\ref{aeq}) and by $b^{1/2}(t)$ in (\ref{beq}),
yielding
\begin{eqnarray}
{d \over  dt} \, a^{1/2}(t) \leq a(t_0) a^{-1/2}(t) + k_{10}
(t - t_0 )^{1/2} + k_{11} b^{1/2}(t) ,
\label{new126}
\end{eqnarray}
\begin{eqnarray}
{d \over  dt} \, b^{1/2}(t) \leq
b(t_0) b^{-1/2}(t) + k_{12}  (t - t_0 )^{1/2} + k_{13} a^{1/2}(t) .
\label{new127}
\end{eqnarray}
We estimate the term $a(t_0) a^{-1/2}(t)$
by using the fact that $a(t)$ is a monotonic increasing function of $t$,
due to positivity of $\hat{\alpha}^2$ in (\ref{a}).
Thus,
$a(t_0) a^{-1/2}(t)$ is bounded by $a^{-1/2}(t_0)$.
In addition, we note the term $k_{10}  (t - t_0 )^{1/2}$
is bounded since $t\in I$ is bounded. We thereby obtain
\begin{eqnarray}
{d \over  dt} \, a^{1/2}(t) \leq k_{14}  + k_{11} b^{1/2}(t) .
\label{aest}
\end{eqnarray}
Similarly, we obtain
\begin{eqnarray}
{d \over  dt} \, b^{1/2}(t) \leq k_{15}  + k_{13} a^{1/2}(t) .
\label{best}
\end{eqnarray}
Adding (\ref{aest}) and (\ref{best}), and defining
$c(t):= a^{1/2}(t)+b^{1/2}(t)$, we derive
\begin{eqnarray}
{d \over  dt} c(t) \leq k_{18} + k_{17} c(t) .
\label{R}
\end{eqnarray}
Gronwall's inequality immediately applies to (\ref{R}),
and so we determine that $c(t)$ is bounded for all $t \in I$.
Then $a^{1/2}(t)$ and $b^{1/2}(t)$, which are positive, are bounded.

Returning to the inequalities (\ref{aeq})-(\ref{beq}),
it follows that ${d \over dt}a(t)$ and ${d\over dt}b(t)$
are each bounded. Hence, from the definitions of $a$ and $b$,
we obtain that $\sup_{\Sigma} \alpha^2$ and $\sup_{\Sigma}\beta^2$
are bounded for all $t\in I$.
Since $\alpha^2 =  (B^a + E^a  )^2$ and $\beta^2 =  (-B^a + E^a  )^2$,
we conclude that $B^a (x,t)$ and $E^a(x,t)$ are bounded
on $\Sigma\times I$.

\subsubsection*{Step 4: Bounded Derivatives}

Now that we have determined that $E^a$, $H^a$, and $B^a$
are bounded on $\Sigma \times I$, we proceed to show that the
first derivatives of these functions,
and subsequently all higher order derivatives,
are bounded on $\Sigma\times I$.

We start with $H^a$.  From (\ref{reducedconstraint}) and
(\ref{reducedEyevolution}), it follows that since $E^a$,
$H^a$, and $B^a$ are bounded, then $\partial_x H^a$ and
$\partial_t H^a$ are bounded. Similarly, if the order $n$
derivatives of
$E^a$, $H^a$, and $B^a$ are bounded,
then it follows from (derivatives of)
(\ref{reducedconstraint}) and (\ref{reducedEyevolution}) that
the order $n + 1$ derivatives of $H^a$ are bounded.
Hence, (by induction on $n$), the derivatives of $H^a$ to all orders
are bounded.

For $E^a$ and $B^a$, we use light cone arguments much like step 3,
but involving a ``derivative stress-energy'' tensor.
Specifically, let
\begin{eqnarray}
T_{(1)\ell \ell} = {1 \over 4}  (\partial_x B + \partial_x E )^2
+ {\lambda \over 4} (v_t + v_n )\partial_x H^a
(\partial_x B^b + \partial_x E^b ) p_{ab}
\label{Tderll}
\end{eqnarray}
\begin{eqnarray}
T_{(1)nn} = {1 \over 4}  (-\partial_x B + \partial_x E )^2
- {\lambda \over 4} (-v_t + v_n )\partial_x H^a
(-\partial_x B^b + \partial_x E^b ) p_{ab}
\label{Tdernn}
\end{eqnarray}
\begin{eqnarray}
T_{(1)\ell n} = -{1 \over 2}  (\partial_x  H )^2
+{\lambda \over 4} (-v_t + v_n )  \partial_x H^a
(\partial_x B^b + \partial_x E^b  ) p_{ab}
\label{Tderln}
\end{eqnarray}
\begin{eqnarray}
T_{(1)n \ell} = -{1 \over 2}  (\partial_x  H )^2
-{\lambda \over 4} (v_t + v_n )  \partial_xH^a
(-\partial_x B^b + \partial_x E^b ) p_{ab}
\label{Tdernl}
\end{eqnarray}
as defined analogously to the stress-energy components
(\ref{Tll}) to (\ref{Tnl}).
Then, as a consequence of the field equations
(\ref{reducedconstraint}) to (\ref{reducedBevolution}),
we find
\begin{eqnarray}
\partial_n T_{(1)\ell \ell} + \partial_{\ell} T_{(1)n \ell} =
Y_1 (E, H, B;  \partial_x E, \partial_x B  )
\label{nderTderll}
\end{eqnarray}
and
\begin{eqnarray}
\partial_\ell T_{(1) nn} + \partial_{n} T_{(1)\ell n} =
Y_2 (E, H, B;  \partial_x E, \partial_x B  )
\label{lderTdernn}
\end{eqnarray}
where $Y_1$ and $Y_2$ are homogeneous quadratic in
$\partial_x E$ and $\partial_x B$, with bounded coefficients.
Although  (\ref{nderTderll}) and (\ref{lderTdernn})
are not strict conservation equations,
we can nevertheless proceed similarly to step 3.

Through use of the field equations,
we can express (\ref{nderTderll}) and (\ref{lderTdernn}) as
\begin{eqnarray}
\partial_n  (\partial_x B + \partial_x E )^2 =
\tilde{Y}_1  (E, H, B; \partial_x E,  \partial_x B )
\label{new137}
\end{eqnarray}
and
\begin{eqnarray}
\partial_{\ell}  (-\partial_x B + \partial_x E )^2 =
\tilde{Y}_2  (E, H, B; \partial_x E,  \partial_x B )
\label{new138}
\end{eqnarray}
with $\tilde{Y}_1$ and $\tilde{Y}_2$ of the same nature as $Y_1$ and $Y_2$.
It then follows from standard algebraic inequalities that
\begin{eqnarray}
\partial_n  (\partial_x B + \partial_x E )^2 \leq
k_{19} | \partial_x B |^2 + k_{20} | \partial_x E |^2 ,
\label{alphader}
\end{eqnarray}
\begin{eqnarray}
\partial_\ell  (-\partial_x B + \partial_x E )^2 \leq
k_{21} | \partial_x B |^2 + k_{22} | \partial_x E |^2 .
\label{betader}
\end{eqnarray}

We now apply the light cone arguments of step 3 to
the differential inequalities (\ref{alphader}) and (\ref{betader}):
Starting at an arbitrary point in $\Sigma \times I$,
we integrate (\ref{alphader}) and (\ref{betader})
back to the initial surface along light rays generated by
$\partial_n$ and $\partial_{\ell}$.
Taking suprema over $\Sigma$ and adding the resulting inequalities,
we obtain
\begin{eqnarray}
&& \sup_{x\in\Sigma} [  (\partial_x B  (x,t) )^2 +
(\partial_x E  (x,t) )^2  ]
\leq  \nonumber \\&&\qquad
k_{23} + k_{24}\int^{t}_{t_0} \sup_{x\in\Sigma}
 [ (\partial_x B  (x,s) )^2 +  (\partial_x E  (x,s)  )^2   ] ds .
\label{Rder}
\end{eqnarray}
Applying the Gronwall inequality to (\ref{Rder})
shows that $ | \partial_x B |$ and $ | \partial_x E |$
are bounded for all $t \in I$.
With $E^a$, $H^a$, $B^a$, $\partial_x E^a$ and $\partial_x B^a$
bounded,  it follows from the evolution equations
(\ref{reducedExevolution}) and (\ref{reducedBevolution})  that
$\partial_t E^a$ and $\partial_t B^a$ are bounded as well.

The previous argument can be applied to all orders of derivatives of
the fields.  This establishes that $E^a$, $H^a$, $B^a$,
and all of their derivatives are bounded on $\Sigma \times I$.
Global existence now follows from the usual ``open-closed''
arguments. \QED

\section{\head{GLOBAL EXISTENCE FOR EQUIVARIANT\\
FRAME FIELD EQUATIONS WITH TORSION}}

As discussed in Section 3,
while left-equivariant wave maps (\ref{leftequivar})
correspond to invariant frame fields (\ref{new57}),
conjugate-equivariant wave maps (\ref{conjequivar})
and right-equivariant wave maps (\ref{rightequivar})
correspond to equivariant frame fields (\ref{new58}).
In this section we show that global existence holds
for smooth solutions to the Cauchy problem for
translation equivariant frame fields.

We first note that by definition of translation equivariance,
$K^a_{\mu}(x, y,t)$ can be expressed as
\begin{eqnarray}
&& K^a_x (x, y, t) = \exp(-yR) E^a (x,t) \exp(yR)
\label{new142}\\
&& K^a_y (x, y, t) = \exp(-yR) H^a (x,t) \exp(yR)
\label{new143}\\
&& K^a_t (x, y, t) = \exp(-yR) B^a (x,t) \exp(yR)
\label{new144}
\end{eqnarray}
in terms of some Lie-algebra valued fields
$\{ E^a(x, t), H^a(x,t), B^a(x, t) \}$
which do not depend on $y$.
Here $R$ is a fixed (constant) element in the Lie algebra;
the left multiplication by $\exp(-yR)$
combined with right multiplication by $\exp(yR)$
denotes the adjoint action of a one-parameter Lie subgroup
on the Lie algebra.

Substituting expressions (\ref{new142}) to (\ref{new144})
into the frame field equations with torsion (\ref{new51}) and (\ref{new54})
on Minkowski space,
we obtain the following 1+1 reduced PDE system
\begin{eqnarray}
\partial _x H^a = &&
- C_{bc}{}^a E^b (H^c -R^c)
\label{new145}\\
\partial _t E^a = &&
\partial _x B^a -C_{bc}{}^a B^b E^c
\label{new146}\\
\partial _t H^a = &&
-C_{bc}{}^a B^b (H^c -R^c)
\label{new147}\\
\partial _t  B^a = &&
\partial _x E^a - C_{bc}{}^a H^b R^c
- C^a{}_{bc}(B^b B^c - E^b E^c -  H^b H^c) \nonumber\\&&\quad
- \lambda Q^a{}_{bc} (v_y B^b E^c - v_x B^b H^c + v_t E^b H^c )
\label{new148}
\end{eqnarray}
provided that $C^a{}_{bc}$ and $Q^a{}_{bc}$ are {\it invariant}
under the adjoint action of $\exp(yR)$.
We note that the only difference
between these equations for translation equivariant frame fields
and equations (\ref{reducedconstraint}) to (\ref{reducedBevolution})
for translation invariant frame fields
is the presence of the commutator terms involving $R$.

While the expressions for the field equations are changed
somewhat in passing from invariant to equivariant frame
fields, the expressions for the stress-energy components
(\ref{Txx}) to (\ref{Tyx}) and (\ref{Tll}) to (\ref{Tnl})
do not change at all.
(In particular, while $K^a_\mu$ is not independent of $y$,
the quadratic expressions
$K^a_\mu K^b_\nu g_{ab}$ and $K^a_\mu K^b_\nu p_{ab}$
are invariant, and consequently so is $T_{\mu \nu}$. )

Initial data for the Cauchy problem for translation equivariant frame fields
consists of Lie-algebra valued functions
$\{{\hat E}^a(x), {\hat H}^a(x), {\hat B}^a (x) \}$
on $\Sigma$ which satisfy the constraint
\begin{eqnarray}
\partial _x {\hat H}^a = - C_{bc}{}^a {\hat E}^b ({\hat H^c} -R^c) . 
\label{new149}
\end{eqnarray}
A solution to the Cauchy problem is a set of fields $\{E^a(x,t), H^a(x,t),
B^a (x,t)
\}$  satisfying equations (\ref{new145}) to (\ref{new148})
and the initial conditions (\ref{initialvalues}).

The global existence result, and its corollary, are stated
as follows.
Let $\Sigma$ denote $R^1$ or $S^1$,
and introduce coordinates $(x,t)$ for $\Sigma\times R^1$.
Fix constants $v_t,v_x,v_y$.
Let $G$ be a Lie group
and let $R^a$ be a fixed (constant) vector in the Lie algebra of $G$.
Assume $G$ admits on its Lie algebra
a positive definite metric tensor $g_{ab}$
and a skew tensor $p_{ab}$
which are each invariant under the adjoint action of
the Lie subgroup generated by $R^a$:
\begin{eqnarray}
&& g_{ae} C_{bc}{}^e R^c = -g_{be} C_{ac}{}^e R^c
\label{ginvariant}\\
&& p_{ae} C_{bc}{}^e R^c = p_{be} C_{ac}{}^e R^c
\label{pinvariant}
\end{eqnarray}
where $C_{bc}{}^a$ denotes the Lie-algebra commutator structure tensor.
Let $Q^a{}_{bc}$ be the tensor defined by (\ref{new50}).

\begin{theorem}
{Theorem 4.}{
Let $\lambda$ be a small constant.${}^4$
For any smooth compact support initial data (\ref{initialvalues})
satisfying (\ref{new149}),
the Cauchy problem (\ref{new145}) to (\ref{new148})
has a unique smooth global solution
$\{E^a(x,t), H^a(x,t), B^a (x,t) \}$ for all $t \in R^1$.}
\end{theorem}

From Propositions~2 and~3 we obtain a corresponding result
for wave maps.

\begin{theorem}
{Theorem 5.}{
The Cauchy problems
for conjugate-translation equivariant wave maps (\ref{conjequivar})
and for right-translation equivariant wave maps (\ref{rightequivar}),
with or without torsion,
have unique smooth global solutions for all smooth compact support
initial data.}
\end{theorem}

\noindent{ \bf Remark 2:}
Under the translation invariant form (\ref{new57}) for frame fields,
which corresponds to left-translation equivariant (\ref{leftequivar}) 
or translation invariant (\ref{transinvar}) wave maps, 
the reduction of the frame field equations
and corresponding wave map equation
is consistent for any Lie group target
with $(G,g,p)$ invariant under left multiplication. 
However, this is not the case
under the translation equivariant form
(\ref{new58}) for frame fields,
which corresponds to conjugate-translation equivariant (\ref{conjequivar}) 
or right-translation equivariant (\ref{leftequivar}) 
wave maps. 
The translation equivariance ansatz
gives a consistent reduction of
the frame field equations and corresponding wave map equation 
only if the target geometry $(G,g,p)$
is invariant under right multiplication by
the translation group generated by the Lie algebra element $R$
appearing in (\ref{leftequivar}) to (\ref{conjequivar}) for wave maps
and (\ref{new58}) for frame fields.
We refer to this condition,
given by (\ref{ginvariant}) and (\ref{pinvariant}),
as {\it translation invariance} of the target.
As shown in Proposition~A in the appendix,
every compact semi-simple Lie group $G$ admits
a translation invariant geometry $(G,g,p)$,
except that the dimension of $G$ must be greater than three
to support a non-zero torsion $Q$ (see Remark~1).

\subsubsection*{Proof of Theorem 4:}
The proof of Theorem 4 is very similar to that of Theorem 2.
We summarize the differences (if any) in each step.

\subsubsection*{Step 1:  Conserved Energy}
Since the expression for the energy is unchanged and since it
is conserved, there are no changes in obtaining $L^2$ bounds
for $E^a(x,t), H^a(x,t), B^a (x,t)$.

\subsubsection*{Step 2:  Bounded $H^a$}
Instead of (\ref{LonexderEy}), we have
\begin{eqnarray}
\int_\Sigma |\partial _x H^a | dx \leq
\int_\Sigma | C_{bc}{}^a E^b H^c| dx
+\int_\Sigma | C_{bc}{}^a E^b R^c | dx .
\label{new150}
\end{eqnarray}
The first of the two terms on the right hand side of
(\ref{new150}) may be handled as in (\ref{ExEyest}).
As for the second term, we have
\begin{eqnarray}
\int_{\Sigma} | C_{bc}{}^a E^b R^c | dx &\leq&
k_{25} \int_{\Sigma} |E^b| dx\nonumber\\
&<& k_{25} \big( \int_{\Sigma} E^2 dx \big)^{1/2}
\nonumber\\
&\leq& k_{26}
\label{new151}
\end{eqnarray}
where the second inequality uses the compact support of $E^b$
together with the Holder inequality,
and the last inequality follows from the $L^2$ bound on $E^a$.
Hence we obtain
\begin{eqnarray}
\int_{\Sigma} | \partial_x H^a| dx \leq
k_{27}
\label{new152}
\end{eqnarray}
analogous to (\ref{LoneEyest}).

If $\Sigma = R^1$, the argument leading to a
pointwise bound on $H^a(x, t)$ for $t \in I$
proceeds exactly as in the proof of Theorem 2.
If $\Sigma = S^1$, then we need to modify the argument
which begins with (\ref{LoneSEy}).
We have
\begin{eqnarray}
\int_{S^1} H^a (x, t) dx &=&
-\int^t_{t_0}\int_{S^1}
C^a{}_{bc} B^b (x, s) \big( H^c (x, s) + R^c \big) dxds\nonumber\\
&& +\int_{S^1} H^a (x, t_0) dx . 
\label{new153}
\end{eqnarray}
The term $\int^t_{t_0}\int_{S^1} C^a{}_{bc} B^b (x, s) R^c dxds$
can be bounded from above using the same
quadratic inequality that is used in (\ref{new151}), and so
we obtain
\begin{eqnarray}
|\int_{S^1} H^a (x, t) dx  | \leq
k_{28}
\label{new154}
\end{eqnarray}
analogous to (\ref{new110}).  The argument for pointwise
bounds on $H^a(x, t)$ for $\Sigma = S^1$ can then be
completed as in the proof of Theorem 2.

\subsubsection*{Step 3:  Bounded $E^a$ and $B^a$}

From inequalities (\ref{nderalpha}) and (\ref{lderbeta}) onward,
the light-cone arguments used to bound $E^a(x, t)$ and $B^a(x, t)$
in the proof of Theorem 2 work identically
to bound $E^a(x, t)$ and $B^a(x, t)$ here.
To arrive at (\ref{nderalpha}) and (\ref{lderbeta}) we use the
following equations, analogous to (\ref{new113}) and (\ref{new114}), 
\begin{eqnarray}
\partial _n (B + E)^2 &=&
-2 H^b ( H^c +R^c ) (B^a + E^a) C_{abc}
\nonumber\\
&& -2 \lambda (v_t + v_x) H^a (- B^b + E^b) (B^c + E^c) Q_{bca}
\nonumber\\
&& +2 \lambda (v_t - v_x) C_{bc}{}^a R^b (B^c + E^c) (B^b - E^b) p_{ab} , 
\label{155}
\end{eqnarray}
\begin{eqnarray}
\partial _n (- B + E)^2 &=&
-2 H^b (H^c +R^c) (- B^a + E^a) C_{abc}
\nonumber\\
&& -2 \lambda (- v_t + v_x) H^a (- B^b + E^b) (B^c +E^c) Q_{bca}
\nonumber\\
&& +2 \lambda (v_t + v_x) C_{bc}{}^a R^b (- B^c+ E^c)( B^b + E^b) p_{ab} . 
\label{156}
\end{eqnarray}
Adopting the notation $\alpha ^a := B^a + E^a$, $\beta^a := -B^a +E^a$,
these equations become
\begin{eqnarray}
\partial _n \alpha^2 &=& -2 C_{abc} H^b (H^c+R^c) \alpha^a
\nonumber\\
&& -2\lambda (v_t + v_x) Q_{bca} H^a \beta^b \alpha^c
+ 2\lambda (v_t + v_x) R^b \alpha^c \beta^b C_{bc}{}^a p_{ab} , 
\label{new157}
\end{eqnarray}
\begin{eqnarray}
\partial _{\ell} \beta^2 &=&
-2 C_{abc} H^b (H^c+R^c) \beta^a
\nonumber\\
&& -2\lambda (-v_t + v_x) Q_{bca} H^a \beta^a \alpha^c
+ 2\lambda (v_t + v_x) R^b \beta^c \alpha^b C_{bc}{}^a p_{ab} . 
\label{new158}
\end{eqnarray}
Then, noting that $C_{abc}$, $Q_{abc}$, $\lambda$, $v_t$, $v_x$ and $R$
are constant,
and recalling that $H^a$ is bounded on $\Sigma \times I$, we obtain
\begin{eqnarray}
\partial _n \alpha^2 \leq k_{29} \sqrt{\alpha^2} + k_{30}
\sqrt{\alpha^2} \sqrt{\beta^2}
\label{new159}
\end{eqnarray}
and
\begin{eqnarray}
\partial _{\ell} \beta^2 \leq k_{31} \sqrt{\beta^2} + k_{32}
\sqrt{\alpha^2} \sqrt{\beta^2}
\label{new160}
\end{eqnarray}
which are identical to (\ref{nderalpha}) and (\ref{lderbeta}).

\subsubsection*{Step 4:  Bounded Derivatives}

One can see in Step 2 and Step 3 that the presence of the
commutator terms involving $R$ in the field equations
(\ref{new145}) to (\ref{new148}) changes little in
the arguments for boundedness, since these extra terms are
easily controlled by the analogous quadratic terms appearing
in the equations.  The same holds true for Step 4.
We can define the derivative stress-energy components just as in
(\ref{Tderll}) to (\ref{Tdernl})
and then obtain conservation equations similar to
(\ref{nderTderll}) to (\ref{lderTdernn}), with small
modifications in the expressions $Y_1$ and $Y_2$ which appear there.
These modifications are readily handled in deriving
the estimate (\ref{alphader}) and (\ref{betader}).
The rest of the argument proceeds unchanged.

Hence we obtain global existence. \QED

\section{\head{CONCLUDING REMARKS}}

The wave map global existence results we have obtained here
extend previous work in two significant ways.
First, our study of translation equivariant critical wave maps
for Lie group targets (Theorems~3 and~5)
provides a counterpart to work on rotationally equivariant critical wave maps
for symmetric-space targets (see \cite{ShatahT-Z,Shatah-Struwe}).
Second, our inclusion of torsion gives an interesting
generalization of critical wave maps for arbitrary targets,
which ties into current work on integrable chiral models in 2+1 dimensions
in the case of Lie group targets \cite{Ward}. 

Furthermore, our results demonstrate the utility of
the frame formulation of wave maps for Lie group targets (Proposition~1).
The translation-equivariant reduction of critical wave maps
studied here is motivated by this formulation
and the analysis is especially straightforward in terms of frames.
An important question to investigate for future work
is how the frame formulation might help in understanding
the unreduced critical wave map equation for general Lie group targets
and symmetric-space targets.

\section*{\head{APPENDIX}}

\begin{theorem}
{Proposition A.}{
Let $G$ be a semi-simple Lie group
with commutator structure tensor $C_{bc}{}^a$.
\begin{enumerate}
\item  The Lie algebra of $G$ admits a
translation invariant (\ref{ginvariant}) positive-definite metric $g_{ab}$
if $G$ is compact.
\item  The Lie algebra of $G$ admits a
translation invariant (\ref{pinvariant}) skew-tensor $p_{ab}$
with non-zero torsion (\ref{new50})
if $G$ is compact and has dimension greater than three.
\item If $G$ has dimension three then the torsion (\ref{new50}) is zero
for every skew-tensor $p_{ab}$ on the Lie algebra of $G$.
\end{enumerate} }
\end{theorem}

\noindent{\bf Proof of 1:}
If $G$ is compact then its Lie algebra admits
an invariant positive-definite metric $g_{ab}$
(see, e.g. \cite{liegroup}), which satisfies
\begin{eqnarray}
g_{ae} C_{bc}{}^e = -g_{be} C_{ac}{}^e .
\end{eqnarray}
(In particular, the Cartan-Killing metric given by
$g_{ab} := -C_{ae}{}^c C_{bc}{}^e$
is  both invariant and positive-definite.)
Hence condition (\ref{ginvariant}) holds.

\noindent{\bf Proof of 2 and 3:}
Hereafter $g_{ab}$ denotes the Cartan-Killing metric.
We first remark that, for any $G$,
the natural construction
\begin{eqnarray}
p_{ab} := C_{ab}{}^d g_{de} R^e
\label{p}
\end{eqnarray}
is easily seen to yield a translation invariant skew-tensor.
But the resulting torsion tensor (\ref{new50}) is always zero, since
\begin{eqnarray}
p_{e[a} C_{bc]}{}^e = C_{e[a}{}^d C_{bc]}{}^e g_{df} R^f =0
\label{zeroQ}
\end{eqnarray}
by the Jacobi identity.

In three dimensions it is easy to show that
$C_{ab}{}^e g_{ec}$ must be proportional to the totally-skew
Levi-Civita tensor $\epsilon_{abc}$,
while any skew-tensor $p_{ab}$
can be expressed in the form
\begin{eqnarray}
p_{ab} = \epsilon_{abc} p^c
\label{p3dim}
\end{eqnarray}
for some vector $p^c$ in the Lie algebra of $G$.
Thus, it follows that $p_{ab}$ must have the form (\ref{p})
where $R^e$ is proportional to $p^e$,
and hence from (\ref{p}) and (\ref{zeroQ})
we have that
the torsion tensor (\ref{new50}) is zero.
This shows that there is no torsion for any three-dimensional $G$
(and hence none in particular with $p_{ab}$ being translation invariant).

Now suppose $G$ has dimension greater than three.
In this case, $G$ must have rank greater than one
and hence the Lie algebra of $G$ possesses an abelian subalgebra
of dimension at least two (see, e.g. \cite{liegroup}).
This allows the explicit construction of a translation invariant
skew-tensor $p_{ab}$ as follows.
Let $p^a$, $q^a$ be any two (linearly independent) commuting vectors 
in the Lie algebra of $G$,
so $p^a q^b C_{ab}{}^c=0$,
and let $p_e:=g_{ea} p^a, q_e:=g_{ea} q^a$.
Set $R^a:= \alpha p^a +\beta q^a \neq 0$ with constants $\alpha,\beta$.
Then it is straightforward to show that the skew-tensor defined by
\begin{eqnarray}
p_{ab} := 2 p_{[a} q_{b]}
\label{skewpq}
\end{eqnarray}
is translation invariant
as a consequence of $p$ and $q$ commuting with $R$.
Now it remains to show that the torsion tensor
given by (\ref{new50}) and (\ref{skewpq}) is non-zero.

We have
\begin{eqnarray}
g_{ad} Q^d{}_{bc} = 3/2( p_e q_{[a} C_{bc]}{}^e - q_e p_{[a} C_{bc]}{}^e ) .
\label{Q'}
\end{eqnarray}
To show that the tensor (\ref{Q'}) is non-zero 
when $G$ is compact,  
we contract (\ref{Q'})
with the vector $s^a = p^a q^e q_e - q^a p^e q_e$
satisfying $s^a q_a=0$.
This yields
\begin{eqnarray}
s^a g_{ad} Q^d{}_{bc} = -1/2 s^a p_a C_{bc}{}^e q_e .
\end{eqnarray}
with $s^a p_a = p^a p_a q^d q_d - (p^a q_a )^2\neq 0$ 
due to positive-definiteness of $g_{ab}$.
Moreover, since $G$ is semi-simple, its Lie algebra has empty center
and so $C_{bc}{}^e q_e = C_{eb}{}^a q^e g_{ca}$ is non-zero
(that is, there exists a vector $v^b$ so that $C_{eb}{}^a q^e v^b \neq 0$). 
Therefore, $s^a g_{ad} Q^d{}_{bc}$ is non-zero
and thus so is the torsion tensor (\ref{Q'}).
\QED

\centerline{\head{ACKNOWLEDGEMENTS}}
\vskip .75cm

Various portions of this work were carried out at the Courant
Institute and at the University of Washington.  
Partial support for this research has come from NSF grant PHY-9800732 
at the University of Oregon.


\end{document}